\documentstyle[preprint,revtex]{aps}	
\begin{document}
%
%
%
\preprint{UH-511-771-93}
\preprint{FSU-HEP-930722}
\preprint{\today}
\begin{title}
ASPECTS OF CHARGINO-NEUTRALINO\\
PRODUCTION AT THE TEVATRON COLLIDER
\end{title}
\author{Howard Baer$^1$, Chung Kao$^1$ and Xerxes Tata$^2$}
\begin{instit}
$^1$Department of Physics,
Florida State University,
Tallahassee, Florida 32306 USA
\end{instit}
\begin{instit}
$^2$Department of Physics and Astronomy,
University of Hawaii,
Honolulu, HI 96822 USA
\end{instit}
\begin{abstract}
We have used ISAJET 7.0/ISASUSY 1.0 to evaluate the rates
and study distributions for trilepton events from the leptonic
decays of charginos and neutralinos produced at the
Fermilab Tevatron via the reaction,
$p\bar p\rightarrow \tilde{W_1} \tilde{Z_2}$ + X
for cuts inspired by the CDF and D0
experiments. We find that about 60\% of these events do not
contain any jets, and so should be easily identifiable over
Standard Model backgrounds from $t\bar{t}$
and WZ production, provided that the decay
$\tilde{Z_2}\rightarrow\tilde{Z_1}Z$ is kinematically inaccessible.
 We further show that, with suitable
cuts, these backgrounds can be reduced to less than 1-2 $fb$  even if
$n_{jet} = 1$, effectively increasing the SUSY sample by 50\%. We
confirm that the signal is only rate-limited, and that the Fermilab Tevatron
after an accumulation of 100 $pb^{-1}$ of integrated luminosity will be able to
explore parameters well beyond the range of LEP and, if sleptons are
substantially lighter than squarks, close to the reach of LEP 200.

\end{abstract}
\pacs{PACS numbers: 14.80.Ly, 11.30.Pb, 13.85.Qk}
\newpage
%
%
\begin{narrowtext}
If low energy supersymmetry (SUSY)\cite{REV} indeed turns out to be nature's
resolution of the fine tuning issue, sparticles should be expected
to reveal themselves as higher energy scales are explored at
colliding beam facilities. Already, the Collider Detector at
Fermilab (CDF) Collaboration\cite{CDF} at the Tevatron, even after
taking into account the cascade decays\cite{CAS}, has excluded squarks
and gluinos lighter than 90-100 GeV, while the experiments at
LEP\cite{LEP} essentially preclude the existence of sleptons and charginos
lighter than $M_Z/2$; the corresponding bounds\cite{LEP} on neutralino masses
are sensitive to the model-dependent mixing angles in the gaugino-Higgsino
sector. These analyses are carried out within the framework of the Minimal
Supersymmetric Model (MSSM) which is the simplest SUSY extension of the
Standard Model (SM) in that it contains the fewest number of new particles and
interactions.

Charginos and neutralinos can also be searched for at hadron
colliders. As shown in Ref.\cite{BT}, the reaction\cite{AN}
$p\bar{p}\rightarrow W^*\rightarrow \tilde{W_1}\tilde{Z_2}$ + X,
$\tilde{W_1}\rightarrow l\nu\tilde{Z_1}$,
$\tilde{Z_2}\rightarrow l'\bar{l'}\tilde{Z_1}$,
which leads
to hadronically quiet, isolated trilepton ($e$ or $\mu$) events for which SM
backgrounds are very small, potentially
provides a promising signal for $\tilde{W_1}\tilde{Z_2}$ production
at the Tevatron.
In a data sample of 100 $pb^{-1}$ in excess of 10-100 trilepton
events are expected\cite{BT,LOPEZ} for
significant
regions of MSSM parameter space beyond the reach of the CERN LEP
collider and, for some ranges of parameters, even beyond the reach of LEP 200.
Thus, the CDF and D0 experiments should be able to probe new
regions of SUSY parameter space by the end of the current Tevatron
run (Run IB). It should, however, be kept in mind that the previous
studies\cite{BT,LOPEZ} on which these conclusions are based
computed only the total trilepton rates without taking into
account experimental cuts and triggers or the hadronic activity
expected from QCD radiation. The main purpose of this paper is
to study these issues in order to better assess the detectability of
the trilepton signal from $\tilde{W_1}\tilde{Z_2}$ production
at the Tevatron. We work within the MSSM framework, and further
assume that the sfermions all have a common mass at the unification
scale as, for instance, in supergravity models.

The production of $\tilde{W_1}\tilde{Z_2}$ pairs at hadron
colliders occurs by $q\bar{q}$ annihilation via $s$-channel $W^*$
exchange or by squark exchange in the $t$ and $u$ channels.
Within the MSSM framework, the production cross section is
fixed in terms of the parameters $m_{\tilde{g}}$, $\mu$,
$\tan\beta$ and $m_{\tilde{q}}$ defined, for instance, in Ref.\cite{BT}.
The cross section for $\tilde{W_1}\tilde{Z_2}$ production at the Tevatron,
computed
using EHLQ\cite{EHLQ} Set I structure functions,
is shown in Fig. 1 versus ({\it a\/}) $m_{\tilde{g}}$,
for $\mu$ = $-$ 200 GeV, and ({\it b\/}) versus $\mu$, for  $m_{\tilde{g}}$ =
300 GeV,
for several values of $m_{\tilde{q}}$. Here, and unless otherwise
specified, in the remainder of
this paper, we have fixed $\tan\beta$ = 2, $m_t$ = 140 GeV, and taken the
pseudoscalar Higgs boson mass, $m_{H_P}$, to be 500 GeV.
The range of $\mu$ between the vertical
lines in Fig. 1{\it b} is excluded by constraints from the LEP
experiments\cite{LEP}
as parameterized in Ref.\cite{BBDKT}. For the $m_{\tilde{q}}$ = $m_{\tilde{g}}$
case in Fig. 1{\it a}, the CDF experiment\cite{CDF} has excluded gluinos
lighter than
about 160-180 GeV; moreover,
the sneutrino is too light unless the gluino is heavier
than about 120 GeV. For the other two cases, the whole range of
$m_{\tilde{g}}$ shown is allowed. Our main purpose in showing this
figure is to demonstrate the effect of the $t$- and $u$-channel
exchange amplitudes, which were not included in the computation of
Ref.\cite{BT}, on the cross section. We see that, if the squark is
relatively light, these contributions can
reduce the total cross section by as much as 40\% due to a negative
interference between the W and $\tilde{q}$ exchange amplitudes.
Nevertheless, this effect rapidly decreases with increasing value of
$m_{\tilde{q}}$; {\it i.e.} the $s$-channel contribution dominates even when
$m_{\tilde{q}}$ = 2$m_{\tilde{g}}$\cite{FN1}.

In order to study the effects of experimental cuts and triggers on
the trilepton signal, we have used the ISAJET Version 7.0/ISASUSY
Version 1.0 program\cite{BPPT} to generate $\tilde{W_1}\tilde{Z_2}$
events in $p\bar{p}$ collisions at $\sqrt{s}$ = 1.8 TeV. The decay
patterns of the charginos and neutralinos as given by the MSSM are
automatically incorporated.
Initial and final state parton showers which are incorporated into
ISAJET\cite{ISAJET} lead to hadronic activity even in those events where the
charginos and neutralinos decay leptonically.
We have modelled the experimental conditions at
the Tevatron by incorporating a toy calorimeter with segmentation
$\Delta\eta\times\Delta\phi =
0.1\times 0.09$ and extending to $|\eta | = 4$ into our simulation. We have
assumed
an energy resolution of $70\% /\sqrt{E_T}$ ($15\% /\sqrt{E_T}$) for the
hadronic (electromagnetic) calorimeter. Jets are defined to be hadron clusters
with $E_T > 15$ GeV in a cone of $\Delta R = \sqrt{\Delta\eta^2 +\Delta\phi^2}
= 0.7$. Leptons are required to be within $|\eta | < 2.5$.

To get some idea of the qualitative features of
the trilepton events, we have studied several distributions for
two different sets of SUSY parameters. For both sets, we
have chosen $m_{\tilde{q}}$ = $m_{\tilde{g}} + 10$ GeV which, in models
with a common sfermion mass at the unification scale leads
to enhanced leptonic decays of $\tilde{Z_2}$, and sometimes, also
of $\tilde{W_1}$. We have
fixed  $\tan\beta$, $m_t$ and $m_{H_P}$ as in Fig. ~1. In Case 1
we have, guided by Fig. 4{\it a} of Ref.\cite{BT}, chosen ``typical''
parameters for which charginos beyond the range of LEP might be searched for at
the Tevatron: here, we have
fixed $m_{\tilde{g}}$ = $-\mu$ = 200 GeV which leads to
$m_{W_1}$ = 77 GeV, $m_{Z_2}$ = 78 GeV, $m_{Z_1}$ = 33 GeV and
$\sigma$(3$l$) = 230$fb$.
In Case 2, we have chosen $m_{\tilde{g}}$ = 350 GeV, $\mu$ = $-$400 GeV;
this leads to a rather heavy chargino neutralino spectrum,
$m_{W_1}$ = $m_{Z_2}$ = 115 GeV and $m_{Z_1}$ = 55 GeV and so
rather different kinematics. For Case 2,
$\sigma$(3$l$) = 37$fb$ so that this is, at best, on
the edge of observability, at least until the main injector becomes
operational.

The transverse momentum distribution of the three hardest leptons in
$\tilde{W_1}\tilde{Z_2}$ events at the Tevatron is shown in Fig.~2{\it a} and
Fig.~2{\it b} for Case~1 and Case~2, respectively. Here, as well as in
those distributions shown in Fig.~3 and Fig.~4, we have not
imposed any cuts on lepton $p_T$, nor have we required that
these leptons be isolated.
As expected, Case~2 leads
to somewhat harder leptons. The dot-dashed curves which show
the $p_T$ distribution of the softest lepton underscore the importance
of being able to detect soft electron and muons in events triggered
by one or two hard leptons.

The jet multiplicities and  $E\llap/_T$ expected in the trilepton
sample from $\tilde{W_1}\tilde{Z_2}$ production are shown by solid
(dashed-dotted) lines for Case 1 (Case 2) in Fig.~3. Although these
events are expected to be free of any hadronic activity at tree-level,
we see from Fig.~3{\it a} that there is at least one jet about 40\% of the
time. We have also checked that for $m_t$ = 120 GeV, only about
15\% of $t\bar{t}$ events  (even fewer for heavier tops)
have $n_{jet} \leq 1$ when both the tops decay semileptonically.
Finally, we see from Fig.~3{\it b} that most of the SUSY events contain
$E\llap/_T$ in excess of  20 GeV.

While hadronically quiet trileptons are indeed a characteristic
feature of $\tilde{W_1}\tilde{Z_2}$ production, we will see later
that there is very little SM background even when these events
contain some jet activity. This led us to investigate the characteristics
of the jets in these events in Fig.~4 where we have shown ({\it a}) the
transverse momentum, and ({\it b}) the pseudorapidity distributions of the
fastest jet
in the events with $n_{jet}\geq 1$ for the two cases discussed
above. Since these jets dominantly come from initial state showering,
it is not surprising to see that the $p_T$($jet$) distribution in
both cases are similar, and backed up against the $p_T$ cut. We
see from Fig.~4{\it b} that these jets are likely to be central
and, only in this sense, resemble those from the decay of a heavy particle.

We now turn to a discussion of the trilepton cross section in
experiments at the Tevatron. For the trilepton events, we further
require:

({\it i}) There are three hard leptons with $p_T$($l_1$) $\geq$ 15 GeV,
$p_T$($l_2$) $\geq$ 10 GeV, and  $p_T$($l_3$) $\geq$ 8 GeV.
We believe that it should be possible to detect isolated electrons
with $p_T$ as small as 8 GeV in events triggered by the two
fastest leptons.
It is also worth noting that it may well be possible to detect
muons with $p_T$($\mu$) as small as 3-5 GeV. In this case, it is
clear from Fig.~2 that the signal will be significantly larger
than our estimate.

({\it ii}) Each lepton is isolated, {\it i.e.}; there is no hadronic activity
exceeding
$0.25p_T(l)$ in a cone with $\Delta R = 0.4$ about the lepton direction. We
have checked that this isolation requirement is more effective in
cutting out the $t$-quark background than requiring the hadronic
activity in the lepton cone to be smaller than 5 GeV. This is because, for
the $t$-quark background,
it is almost always the lepton from the $b$-quark that is the softest
and least likely to be isolated. For $p_T$($l_3$) smaller than 20 GeV,
our isolation requirement is clearly the more stringent, and hence, more
effective in removing the background, with little loss of signal.

({\it iii}) The jet multiplicity, $n_{jet}\leq 1$. As discussed above, this
retains about 85\% of the signal and only a small fraction of the top
background. We see from Fig.~3{\it a}, about 2/3 of these events will be
``gold-plated'' in that they are free of jet activity. Events with multiple
jets can be confused for top-quark or squark or gluino production\cite{BKT}.

({\it iv}) We veto events where any pair of oppositely charged leptons
reconstructs the Z mass within $\pm 10$ GeV\cite{FN2}. This reduces the signal
by about 4\% (8\%) for Case 1 (Case 2) but should essentially eliminate most of
the WZ background.

We have checked that with these cuts, the experimental efficiency
for the trilepton signal is 35\% (46\%) for Case 1 (Case 2).
The efficiency is mainly governed by the masses of the parents
and the mass difference between them and the LSP. As expected,
it is smallest for lighter charginos and neutralinos where the
production cross section is the largest, and generally increases with
increasing values of $\tilde{W_1}$ and $\tilde{Z_2}$ masses. It
should be kept in mind that the actual efficiency will
sensitively depend (see Fig. 2) on the capabilities of the detectors
to identify a third lepton with low $p_T$
in events triggered by one or two harder leptons.

The Tevatron cross sections for the trilepton signals, with these cuts,
is shown in Tables I and II for $m_{\tilde{q}}$ = $m_{\tilde{g}} + 10$ GeV
and  $m_{\tilde{q}}$ = 2$m_{\tilde{g}}$, with ({\it a}) $\tan\beta$ = 2
and ({\it b}) $\tan\beta$ = 20. The dashes in the Tables denote
parameter values already excluded by experiments at LEP\cite{LEP,BBDKT}.
For each parameter set in these tables, we have generated 5K
$p\bar p\rightarrow \tilde{W_1} \tilde{Z_2}$ + X events, and forced
the leptonic decay of both the $\tilde{W_1}$ and the $\tilde{Z_2}$.
Thus even for an acceptance of just 10\% after cuts, we see that
the statistical error on the cross sections is small.
We see from these Tables that after
the cuts, the trilepton cross sections are rather small for parameter
values allowed by experiment. Nevertheless, up to 20 trilepton events
may be expected in the CDF and D0 experiments, assuming that
each experiment accumulates, as expected, an integrated
luminosity of about 50 $pb^{-1}$ during the course of the current Tevatron run.
We will see below that SM backgrounds to the trilepton sample are very small
so that even a handful of events could signal the existence of new physics.
The following features of these tables are worthy of note.
\begin{itemize}

\item
The cross sections are largest when squarks and gluinos are close in mass.
In this case, sleptons are considerably lighter than squarks.
As discussed in Ref.\cite{BT}, this leads to an enhancement of the
leptonic decays of $\tilde{Z_2}$, and sometimes, also of $\tilde{W_1}$
resulting in a corresponding increase of the trilepton cross section.

\item
The signals tend to decrease with increasing values of $\tan\beta$.
While this is mainly due to the variation of the leptonic
decays discussed above\cite{BT}, some variation also comes from a change in
the $\tilde{W_1}\tilde{Z_2}$ production rate.

\item
For the favourable case when $m_{\tilde{q}} \approx m_{\tilde{g}}$, and
small values of $\tan\beta$, the CDF and D0 experiments should be
able to probe gluino masses up to about 300 GeV at the end of the
current Tevatron run. Even for the $\tan\beta = 20$ case in Table I,
we see that several trilepton events are expected for
$m_{\tilde{g}} = 300$~GeV provided $|\mu|$
is large. This is entirely reasonable if we recall that for
large values of $|\mu|$, the lighter neutralinos become gaugino-like
and decouple from the Z, increasing the importance of the sfermion-mediated
decays of $\tilde{Z_2}$; since sleptons are considerably lighter than
squarks, the leptonic decays dominate.

\item
For the case of heavy squarks in Table II, we see that it should be
possible to probe gluino masses up to about 200-250 GeV, given
a Tevatron data sample of 100 $pb^{-1}$. Notice that the signal is
very small for the $\tan\beta = 20$ case in Table~IIb. This is because
of a sharp drop in the leptonic decays of $\tilde{Z_2}$.

\end{itemize}

We remind the reader that roughly 60\% of these trilepton events will
not contain any jets, and so should be very distinctive.
It should also be kept in mind that the cross sections shown in the Tables
are sensitive to the experimental cuts. In particular, significantly
larger rates may be obtained if it is experimentally feasible
to identify events where $p_T$($l_3$) is smaller than 8 GeV. Since
this is likely to be the case at least for muons, the rates shown in the Tables
may be regarded as conservative.

The dominant SM backgrounds to the trilepton signal come from ({\it i\/})
$W^{\pm}Z$ production, where both the W and Z decay leptonically, and ({\it
ii\/}) from $t\bar{t}$ production where both tops decay leptonically and a
lepton
from one of the daughter b-quarks is accidently isolated\cite{FN3}. In order
to estimate these backgrounds we have generated 10K $W^{\pm}Z$ and
10K $t\bar{t}$ events, and forced the leptonic decays of the W and Z
bosons as well as of the top quark. The
cross sections for trileptons from these SM background sources with the
same cuts as for the signal is shown in Table III for three values
of the $t$-quark mass. Trileptons from $W^\pm Z$ production should be
easily distinguishable from the signal since these events contain a
dilepton pair that reconstructs the Z mass within experimental resolution
(note that we have not applied the cut ({\it iv}) vetoing these events
for the $W^\pm Z$ background in Table III). We see that about
1.5 trilepton events from this source are expected per 100 $pb^{-1}$ at
the Tevatron. These events should be readily distinguishable from
the signal (unless the decay $\tilde{Z_2} \rightarrow Z\tilde{Z_1}$ is
kinematically accessible; we have checked that this decay mode is
closed for the parameters in the Tables) since one pair of leptons
must reconstruct to the Z mass within experimental resolution.
The mass cut ({\it iv}), already
implemented in Tables I and II, should eliminate most of the $W^\pm Z$
background. We have also checked that applying a more stringent cut vetoing
events within 15 GeV of $M_Z$ reduces the cross sections in Tables I
and II by less than about 5\%. Finally, we see from Table III that top quark
production leads to less than 0.01-0.2 trilepton events per 100 $pb^{-1}$
at the Tevatron, and so, is not a significant background. We thus
conclude that with the cuts ({\it i-iv\/}) discussed above, the trilepton
signal from neutralino-chargino production at the Tevatron is essentially
rate-limited. With an integrated luminosity of 100 $pb^{-1}$, experiments at
the Tevatron should be able to probe gluino masses up to 300 GeV for the
favourable case $m_{\tilde{q}}\approx m_{\tilde{g}}$, and up to 200-250 GeV if
squarks are heavy.

In Fig. 5, we have illustrated the  $M_{l\bar{l}}$ distribution in the signal
events for the two cases discussed earlier. We have included both
the oppositely charged lepton pairs in each event in the figure.
The mass distribution of the lepton pair originating from the
decay of $\tilde{Z_2}$ cuts off at $m_{\tilde{Z_2}} - m_{\tilde{Z_1}}$
while the other pair of leptons is expected to exhibit a broader mass
distribution. This accounts for the kink at
$M_{l\bar{l}} = m_{\tilde{Z_2}} - m_{\tilde{Z_1}}$ in Fig. 5. A study
of this distribution could, therefore, yield information about
the neutralino masses. Accumulation of enough events for this to be
practical could, however,
take several years of Tevatron operation depending on model parameters.

To summarize, we have, within the MSSM framework, examined the trilepton signal
from $\tilde{W_1}\tilde{Z_2}$ production at the Tevatron, taking into account
experimental cuts. We find that the leptonic decays of the chargino and
neutralino lead to spectacular trilepton events at the Tevatron. Typically,
60\% of these events are free from hadronic activity while the bulk of the
remaining events contain a single jet. The general features of these events are
summarized in Fig. 2-4 for representative values of parameters. We have shown
that the SM backgrounds (see Table~III) can be reduced to insignificant levels
by a simple set of cuts.
The signal cross sections, after these cuts are summarized in Tables I
and II. Assuming an integrated luminosity of 100 $pb^{-1}$ that is expected to
be accumulated by the end of the current run of the Tevatron, the CDF and D0
experiments should collectively be sensitive to gluino masses up to 300 GeV if
$m_{\tilde{q}} = m_{\tilde{g}}$ whereas, for heavy squarks, gluinos with masses
up to 200-250 GeV might be detectable. Thus, this trilepton signal can probe
regions of the MSSM parameter space not accessible at LEP or via the direct
search for gluinos at the Tevatron. Finally, we note that with a data sample of
1 $fb^{-1}$ that should be accumulated at the Tevatron if the main injector
becomes operational\cite{FN4}, we see that it may be possible to
detect gluinos as heavy as
400 GeV for some ranges of SUSY parameters assuming, of course, that it is
possible to veto $Z\rightarrow l\bar{l}$ decays with a high efficiency.

%
\acknowledgements
One of us (HB) is grateful to the RCMP for hospitality during the
course of this work.
This research was supported in part by the U.~S. Department of Energy under
contract number DE-FG05-87ER40319 and DE-AM03-76SF00235.
\bigskip

\newpage
%
%
%
%

%
\newpage
%
%
\figure{Total cross sections for $\tilde{W_1}\tilde{Z_2}$ production
in $p\bar p$ collisions at $\sqrt s=1.8$~TeV,
versus ({\it a\/})$m_{\tilde g}$ for $\mu$ = $-$200 GeV, and ({\it b\/}) $\mu$,
for
$m_{\tilde{g}}$ = 300 GeV. We have fixed $\tan\beta =2$, $m_t =140$ GeV,
$m_{H_p}=500$ GeV and shown the cross sections for $m_{\tilde q}=m_{\tilde g}$
(solid), $m_{\tilde q}=2m_{\tilde g}$ (dashed), and $m_{\tilde q}=4m_{\tilde
g}$ (dashed-dotted). The region between the vertical lines
in Fig.~1{\it b} is excluded by the LEP data as discussed in the text.
\label{FIG1}}
%
\figure {The $p_T$ distributions of the three leptons in
trilepton events from chargino-neutralino production at the Tevatron
for ({\it a\/}) Case~1, and ({\it b\/}) Case~2 introduced in the text.
\label{FIG2}}
%
\figure{({\it a\/}) The jet multiplicity, and ({\it b\/}) $E\llap/_T$
distributions
in trilepton events from $\tilde{W_1}\tilde{Z_2}$ production at
the Tevatron for Case~1 (solid) and Case~2 (dashed-dotted) discussed
in the text.
\label{Fig3}}
%
\figure{({\it a\/}) The transverse momentum and ({\it b\/}) pseudorapidity
distributions
of the fastest jet
in trilepton events with $n_{jet}\geq 1$
 from $\tilde{W_1}\tilde{Z_2}$ production at
the Tevatron for Case~1 (solid) and Case~2 (dashed-dotted) discussed
in the text.
\label{Fig4}}
\figure{ The mass distribution of all unlike sign dilepton pairs (two per
event) in the trilepton signal for Case~1 (solid) and Case~2 (dashed-dotted)
after the cuts ({\it i\/})-({\it iii\/}) discussed in the text. The kink in the
distribution occurs at $m_{\tilde{Z_2}}-m_{\tilde{Z_1}}$.
\label{Fig5}}
\eject
%

\begin{table}
\caption{Signal cross sections in {\it fb\/} at the Tevatron after cuts
discussed in the text for $m_{\tilde{q}} = m_{\tilde{g}} + 10$ GeV.
The dashes denote parameter values excluded by LEP experiments.}

\begin{tabular}{ccccccccc}
$m_{\tilde g}$/ $\mu$ &
$-400$ & $-300$ & $-200$ & $-100$ & 100 & 200 & 300 & 400 \\
\tableline
{\bf (a) $\tan \beta= 2 $ }   & & & & & & & & \\
150 & 260 & 230 & 160 &  80   & - & -   & -  & -  \\
200 & 120 & 100 &  82 &  21   & - & -   & -  & -  \\
250 & 60  &  53 &  39 &  2.7  & - & -   & -  & 65 \\
300 & 32  &  26 &  17 &  0.45 & - & -   & 39 & 34 \\
400 & 8.8 & 6.3 & 1.5 &  0.26 & - & 5.8 & 11 & 11 \\
{\bf (b) $\tan \beta= 20$ }  & & & & & & & & \\
150 & -   & -   & -    & -   & -   & -     & -   & -   \\
200 & 140 & 130 & -    & -   & -   & -     & -   & -   \\
250 & 97  & 74  & 20   & -   & -   & -     & 48  & 74  \\
300 & 55  & 31  & 4.6  & -   & -   & 4.1   & 24  & 40  \\
400 & 10  & 2.9 & 0.15 & 1.2 & 1.3 & 0.034 & 3.3 & 8.6
\end{tabular}
\end{table}

\medskip

\begin{table}
\caption{The same as Table I except that $m_{\tilde{q}} = 2m_{\tilde{g}}$. }

\begin{tabular}{ccccccccc}
$m_{\tilde g}$/ $\mu$ &
$-400$ & $-300$ & $-200$ & $-100$ & 100 & 200 & 300 & 400 \\
\tableline
{\bf (a) $\tan \beta= 2 $ } & & & & & & & & \\
150 & 190 & 160 & 84  & 13  & - & -   & -   & -    \\
200 & 75  & 45  & 16  & 7.1 & - & -   & -   & -    \\
250 & 24  & 11  & 5.7 & 4.4 & - & -   & -   & 0.97 \\
300 & 6.5 & 3.6 & 3.5 & 2.8 & - & -   & 3.9 & 0.32 \\
400 & 1.2 & 1.4 & 1.6 & 1.0 & - & 2.6 & 3.3 & 1.7  \\
{\bf (b) $\tan \beta= 20$ } & & & & & & & & \\
150 & -   & -   & -   & -   & -   & -   & -   & -   \\
200 & 1.6 & 3.0 & -   & -   & -   & -   & -   & 12  \\
250 & 1.9 & 3.9 & 8.1 & -   & -   & -   & 5.4 & 3.9 \\
300 & 2.0 & 3.9 & 6.8 & -   & -   & 7.9 & 5.2 & 3.1 \\
400 & 1.6 & 2.3 & 3.3 & 3.0 & 2.4 & 3.9 & 2.8 & 2.0
\end{tabular}
\end{table}

\newpage

\begin{table}
\caption{Cross sections in {\it fb\/} of the backgrounds from $W^{\pm}Z$
and $t\bar{t}$ produced at Tevatron after cuts. The $W^{\pm}Z$ background was
evaluated without the $M_{l\bar{l}}$ cut ({\it iv\/}) introduced in the text. }

\begin{tabular}{cccc}
$W^{\pm}Z$ & $t\bar{t}$, $m_t =$ 120 GeV & $t\bar{t}$, $m_t =$ 140 GeV
& $t\bar{t}$, $m_t =$ 160 GeV \\
\tableline
15  & 1.7 & 0.55 & 0.15
\end{tabular}
\end{table}
%
\end{narrowtext}

\begin{references}
\bibitem{FN2} The D0 experiment will, of course, not be able to
veto $e^+e^-$ pairs since they do not have a central magnetic
field. It should, however, be straightforward to check the signal
loss if events with any electron pair in the mass range are vetoed.
\bibitem{FN4} Although the main injector is not expected to become
operational until well after LEP 200, the trilepton signal should
still be of interest for direct detection of $\tilde{Z_2}$ since the production
of neutralinos at LEP 200 may be very small if the selectron is heavy and the
$|\mu|$ is large.
\end{references}
\end{document}